\documentclass{emulateapj}

\usepackage{amssymb}
\usepackage{graphicx}

\newcommand{\be}{\begin{equation}}
\newcommand{\en}{\end{equation}}
\newcommand{\ro}{R_{\circ}}
\def\etal{{\it et al}. }
\def\msun{\rm M_{\odot}}

\def\msunyr{\rm M_{\odot}\,yr^{-1}}

\def\kms{\rm \, km \, s^{-1}}
\def\gcm2{\rm g\,cm^{-2}}
\def\gcm3{\rm g\,cm^{-3}}
\def\cm3{\rm \, cm^{-3}}

\begin{document}

\title{Gravitational Collapse and Filament Formation: Comparison with the Pipe Nebula}

\author{Fabian Heitsch\altaffilmark{1,2}}
\author{Javier Ballesteros-Paredes\altaffilmark{3}}
\author{Lee Hartmann\altaffilmark{2}}
\altaffiltext{1}{Dept. of Physics \& Astronomy, University of North Carolina at Chapel Hill,
CB 3255, Chapel Hill, NC 27599-3255, U.S.A}
\altaffiltext{2}{Dept. of Astronomy, University of Michigan, 500 Church
St., Ann Arbor, MI 48109-1042, U.S.A}
\altaffiltext{3}{Centro de Radioastronom\'{i}a y Astrof\'{i}sica, UNAM, Apdo. Postal 72-3 
                 (Xangari), Morelia, Michoac\'{a}n 58089, Mexico}
\lefthead{Heitsch et al.}
\shorttitle{Physics of the Pipe Nebula}
\shortauthors{Heitsch et al.}

\begin{abstract}
Recent models of molecular cloud formation and evolution suggest that
such clouds are dynamic and generally exhibit gravitational collapse.
We present a simple analytic model of global collapse onto a
filament and compare this with our numerical simulations 
of the flow-driven formation of an isolated molecular cloud
to illustrate the
supersonic motions and infall ram pressures expected in models of
gravity-driven cloud evolution.  We
apply our results to observations of the Pipe Nebula, an
especially suitable object for our purposes as its low star
formation activity implies insignifcant perturbations from stellar
feedback.  We show that our collapsing cloud model can explain the
magnitude of the velocity dispersions seen in the $^{13}$CO
filamentary structure by Onishi et al.  and the ram pressures required
by Lada \etal to confine the lower-mass cores in the Pipe nebula.  We
further conjecture that higher-resolution simulations will show small
velocity dispersions in the densest core gas, as observed, but which
are infall motions and not supporting turbulence.  Our results point
out the inevitability of ram pressures as boundary conditions for
molecular cloud filaments, and the possibility that especially
lower-mass cores still can be accreting mass at significant rates, as
suggested by observations.
\end{abstract}


\section{Introduction}

The supersonic, ``turbulent'' motions observed in molecular clouds
must play an important role in star formation.  Early numerical models
of molecular clouds often imposed supersonic velocities as either a
continuous forcing term 
\citep{1998PhRvL..80.2754M,1998ApJ...508L..99S,1999ApJ...526..279P}
and/or as initial conditions \citep{2002MNRAS.332L..65B,2003MNRAS.339..577B}.
Yet without an understanding of how supersonic
turbulence originates, it is difficult to develop a predictive theory
of the processes leading to the formation of stars.

The recognition that molecular clouds might often if
not generally result from accumulation of gas by large scale flows
\citep{1999ApJ...527..285B,1999ApJ...515..286B,2001ApJ...562..852H} 
and in particular from atomic flows
(\citealp{2002ApJ...564L..97K}; \citealp{2005A&A...433....1A};
\citealp{2006ApJ...648.1052H}) has made it plausible that 
the turbulence is arising as a consequence of the
cloud's formation.  \citet{2007ApJ...657..870V} showed that turbulence could
develop in clouds formed by variable-velocity flows.  By imposing a
fixed spatial variation in the supersonic inflow velocities, they in
effect identified the driving mechanism as variations in the inflow
speeds (also \citealp{2008A&A...486L..43H}).  In contrast,
\citet{2006ApJ...648.1052H,2008ApJ...674..316H} showed that even uniform
inflows can produce turbulent substructure if the shock interface is
not planar and/or precisely perpendicular to the flows.

It is clear from the studies described in the previous paragraph
that molecular clouds swept up by supersonic flows in the interstellar
medium will begin their existence as both structured and turbulent.
However, this sweep-up by itself only results in cold clouds that are at most 
mildly supersonic if not subsonic 
\citep{2002ApJ...564L..97K,2005A&A...433....1A,2006ApJ...648.1052H,2008ApJ...674..316H}.
The dominant mechanism for producing {\em supersonic} "turbulent" motions in molecular clouds -- especially at
column densities typical of star-forming clouds -- is gravity
(\citealp{2007ApJ...657..870V,2008MNRAS.390..769V,2009arXiv0904.4515V};
\citealp{2008ApJ...674..316H,2008ApJ...689..290H}).
This was already seen by 
\citet{2004ApJ...616..288B}, who
pointed out that clouds with many Jeans masses and non-spherical
geometry are generically susceptible to generating large,
spatially-variable gravitationally-driven flows,
as commonly seen in simulations with non-periodic gravity allowing
global collapse
\citep{2002MNRAS.332L..65B,2003MNRAS.339..577B,2008ApJ...674..316H,2007ApJ...657..870V,2009arXiv0904.4515V}.
\citet{2007ApJ...654..988H} went
further to argue that large-scale gravitational collapse is a feature
of at least the Orion~A molecular cloud \citep[see also][]{2009ApJ...697.1103T}.

Relatively quiescent regions, undisturbed by energy input
from young stars, can provide good tests of the
gravity-driven picture of cloud evolution.
The Pipe Nebula is a prominent and well-studied
example of a cloud of significant
molecular gas mass with little perturbing star formation
\citep{2006A&A...454..781L,2009arXiv0908.4086F}. The densest regions
of the Pipe generally lie along a well-defined filamentary structure
\citep{2007A&A...462L..17A,2007ApJ...662.1082R,2007ApJ...671.1820M,2008ApJ...672..410L,2009arXiv0904.4169R}
-- a common feature of star-forming clouds \citep[e.g.,][]{1979ApJS...41...87S}.
While much attention has been paid by the above authors
to the mass function of the dense cloud cores found along the filaments,
our interest is in the dynamical environment and evolution of these cores.

\citet{2008ApJ...672..410L} concluded that significant external
pressures were needed to confine many of Pipe cores, and
attributed these pressures to "the weight of the surrounding molecular cloud".
As the surrounding cloud is highly unlikely to be in hydrostatic equilibrium, this
"weight" is most likely to be a dynamic pressure.  In particular,
as the filament and cores represent a significant mass concentration,
gravitational acceleration of external material should provide a
confining pressure.  A signature of this confining infall should
be supersonic line broadening.  While the internal non-thermal
velocity dispersions of the Pipe cores are subsonic, as detected
in dense gas tracers \citep[see also][for other regions]{1983ApJ...270..105M,
2007A&A...472..519A,2007ApJ...668.1042K}, \citet{1999PASJ...51..871O} found supersonic line
widths in $^{13}$CO, qualitatively what would be expected from
gravitationally-accelerated, infalling material which enters the
cores and filaments through shocks at the boundaries
\citep{2007ApJ...669.1042G,2009ApJ...699..230G}. Our goal
here is to make this qualitative picture more quantitative.

In this paper we attempt to develop an understanding of the
gas flows driven by global gravitational collapse onto a dense 
filament. We start out with an analytical model to estimate the
expected velocities in the collapsing gas (\S\ref{s:toymodel}). 
This model is in turn motivated by numerical simulations 
(\S\ref{s:numresults}) of the flow-driven formation of an
isolated low-mass cloud. These simulations show that filament
formation and global collapse are a natural consequence of the cloud
formation process. We compare the analytical and numerical results,
and then use them as a guide to interpret recent observations of dense
gas filaments in the Pipe Nebula (\S\ref{s:pipecomp}).
We find that the typical velocity
dispersions resulting from collapse of gas onto the dense filament 
agree reasonably well with
observations of gas at the densities which we probe in the simulation.
We also find that the ram pressures in the numerical simulation agree
reasonably well with the confining pressures inferred by
\citet{2008ApJ...672..410L} for their lower-density cloud cores.  Our models
imply that the observed non-thermal velocity dispersions are likely 
due to infall rather than supporting turbulence, and that the lower-mass
cores can be gaining mass at significant rates.  We conclude that the
gravity-driven picture of cloud evolution is in reasonable agreement
with the observations of this quiescent molecular cloud region.
 
\section{The infinite cylinder and infall}\label{s:toymodel}

Before examining the results from the numerical simulations, it is
useful to develop estimates from a simple model to serve as a
benchmark.  Consider a uniform, infinite, self-gravitating filament
extended in the $z$ direction, with $R$ the radial distance in
cylindrical coordinates.  The critical line density $m_c$ (mass per
unit length in the $z$ direction), obtained by integrating momentum
balance from $R=0$ to $R = \infty$ is a function only of temperature
\citep{1964ApJ...140.1056O}; 
\be 
m_c ~=~ 2 c_s^2/G\,, 
\en 
where $c_s$ is the isothermal sound speed.
Assuming a mean molecular weight of $2.36 m_H$,
\be 
m_c ~=~ 16.3 \, T_{10}\, \msun \,{\rm pc}^{-1}\,, \label{eq:linecrit} 
\en
where $T_{10}$ is the gas
temperature in (typical molecular cloud) units of 10~K.  The filament
has a density structure as a function of cylindrical radius of 
\be
\rho ~=~ \rho_0 \, ( 1 ~+~ R^2/(4 H^2))^{-2}\,, 
\en 
where the scale height $H$ is given by
\be 
H ~=~ c_s^2/(2 G \Sigma_0) ~=~ 0.19 \, T_{10} \, A_V^{-1}\, {\rm pc} \,. \label{eq:h}
\en
Here we have assumed that the relation between extinction and molecular hydrogen
column density is $A_V = 1.1 \times 10^{21}/N(H_2)$.

We may also relate the scale height to the central density;
\be
H^2 ~=~ {c_s^2 \over 2 \pi G \rho_0}\,. \label{eq:hrho}
\en
For reference, the half-mass radius is at $R = 2H$.

Now consider a parcel of gas in free-fall toward this cylinder,
starting from rest at a cylindrical radial distance $\ro$.  The
gravitational acceleration is
\be
a_G = - 2 G m/R\,,
\en
and thus the infall velocity $v$ at $R$ is 
\be
v = 2 (G m \ln (\ro/R))^{1/2} \,. \label{eq:vm}
\en
If we put this in terms of a static filament,
\begin{eqnarray}
v&=& 2^{3/2} c_s (m/m_c)^{1/2}) (\ln (\ro/R))^{1/2} \\\nonumber &=&
0.53 \ln (\ro/R))^{1/2} (m/m_c)^{1/2}\, T_{10}^{1/2}
\kms\,. \label{eq:vin}
\end{eqnarray}
The infall velocity is therefore relatively insensitive to the initial
position.

In a realistic situation, infalling material will shock at the
filament boundary as it adds mass to the filament.  The subsequent
velocity dispersion in the post-shock, filamentary gas will be
subsonic but non-zero 
\citep{2007ApJ...669.1042G,2009ApJ...699..230G}, 
with a precise value depending upon the rate of
radiative cooling.

This model ignores any global motions which are generally present in
finite clouds.  \citet{2004ApJ...616..288B} showed that the collapse of
a circular, uniform sheet results in a strong pile-up of material at
the infalling edge ("gravitational edge focusing")\footnote{We should
point out that this effect has also been noted by \citet{2001ApJ...556..813L},
who used it to explain clustered star formation, albeit in the context of
magnetically dominated finite sheets subject to ambipolar drift.}.
Interior to this
edge, the collapse timescale
$t_c$ of a region of extent $\delta r$ is relatively independent of
radial position $r$, and is approximately 
\be 
t_c \sim \left({ R \over
  \pi G \Sigma} \right)^{1/2}\\, \label{eq:tc} 
\en 
where $R$ is the
inital radius of the sheet and $\Sigma$ is the surface density.  This
is also approximately the time taken for the edge of the sheet to
reach the center.  Equation (\ref{eq:tc}) implies that the velocity
difference across a region $\delta r$ is roughly 
\be \delta v \sim \left({ \pi G \Sigma \over R} \right)^{1/2} \delta
r\,. \label{eq:deltav} \en
The relative importance of the global collapse to the filament-induced
velocities will then depend upon relative magnitudes of the surface
density of the external region and the mass line density of the
filament.

\section{Numerical results}\label{s:numresults}

With the analytical estimates of the flow dynamics
around filaments in hand, we apply them now to a less 
idealized geometry,
using a simulation of cloud growth driven by large-scale
gas flows. The goal is to
test whether the analytical estimate can reproduce the
qualitative behavior of gas collapsing onto a dense filament in a more
complex environment. We briefly summarize the main
properties of the simulation (for further details, see
\citealp{2008ApJ...674..316H}). The simulation models the
flow-driven formation of an isolated molecular cloud,
including the appropriate heating and cooling processes
as well as self-gravity. Two gas flows colliding head-on
at a shocked interface lead to compression, strong cooling
and rapid fragmentation due to a combination of dynamical
and thermal effects. The isolated cloud forming out of this
collision is mildly turbulent, and with increasing mass
it submits to global gravitational collapse perpendicularly
to the inflows. The finite cloud geometry leads to a sweep-up
of material due to global gravitational edge focusing 
(see \citealp{2004ApJ...616..288B}), resulting in the formation
of a dense filament at the cloud {\em edge} (see Fig. 1 of 
\citealp{2008ApJ...689..290H}). As we will show below,
the gas infall onto the filament itself is roughly cylindrical.
It is the infall of gas onto this filament which we are interested in.

This formation mechanism is able to
reproduce some of the salient properties of molecular clouds,
namely their internal turbulence, the predominantly filamentary
structure of their dense gas, and the observed rapid onset of
"star" formation in the clouds. 
The setup is rather generic and is physically equivalent
to e.g. the collision of two supernova shells, the sweep-up
of gas by an expanding shell \citep[e.g.,][]{1998ApJ...507..241P}, or gas swept
up in spiral arms of galaxies 
\citep{1979ApJ...231..372E,2007ApJ...668.1064E,2003ApJ...599.1157K,2007MNRAS.376.1747D}.

With a box size
of $22\times 44^2$~pc and a resolution of $256\times 512^2$ cells, 
our simulation does not
have the spatial resolution to directly model the small cores seen in
the Pipe, since it does not reach the required densities and temperatures.
Yet, we can use it to check the approximate validity of the
simplistic model of infall onto a filament described in the previous
section under more realistic conditions, including the cloud's evolution and
global gravitational collapse.

To compare the gas infall in our simulation to the physically appropriate
tracer, we estimate the $^{13}$CO line emission from the model cloud.
For this estimate, we make a simple approximation, motivated by the notion that CO formation
requires shielding by dust grains (see \citealp{2008ApJ...689..290H} for more
details).
We decide whether CO is ``present'' in a particular grid cell by
determining the attenuation of the ambient radiation field integrated over solid angles.
If the effective extinction is equivalent to that of an
angle-averaged $A_V= 1$ and the local temperature is $T<50$~K, we
assume CO is present in high abundance.  Because CO is rapidly
dissociated at lower extinctions \citep{1988ApJ...334..771V}, we do
not advect CO for simplicity.  The radiation field at each grid point
is calculated by measuring the incident radiation for a given number
of rays and averaging over the resulting sky.
 The ray number is determined such that at a radius corresponding to
$n_c=256$ cells (i.e. half the size of the larger box dimensions),
each resolution element of the cartesian model grid is hit by one ray, i.e. 
$n_{ray} = 4\pi n_c^2$. Thus,
fine structures and strong density variations are resolved (see also
\citealp{2006MNRAS.373.1379H}).  With the distribution of
CO thus determined, we apply a three-dimensional version of the
radiative transfer Montecarlo code by \citet{1979A&A....73...67B}, as implemented
by Mardones et al. (in preparation). The code
determines the level populations by assuming statistical equilibrium
between collisions and radiation. The collision rates are taken from
the Leiden Atomic and Molecular Database\footnote{\tt http://www.strw.leidenuniv.nl/$\sim$moldata/}. 
To lower the computational
needs, we reduce the resolution of the original data cube by a factor of $2$.
After few tens of iterations,
the level populations converge, and the line profiles in an arbitrary
direction can be obtained.

\begin{figure*}
  \begin{center}
    \includegraphics[width=0.8\textwidth]{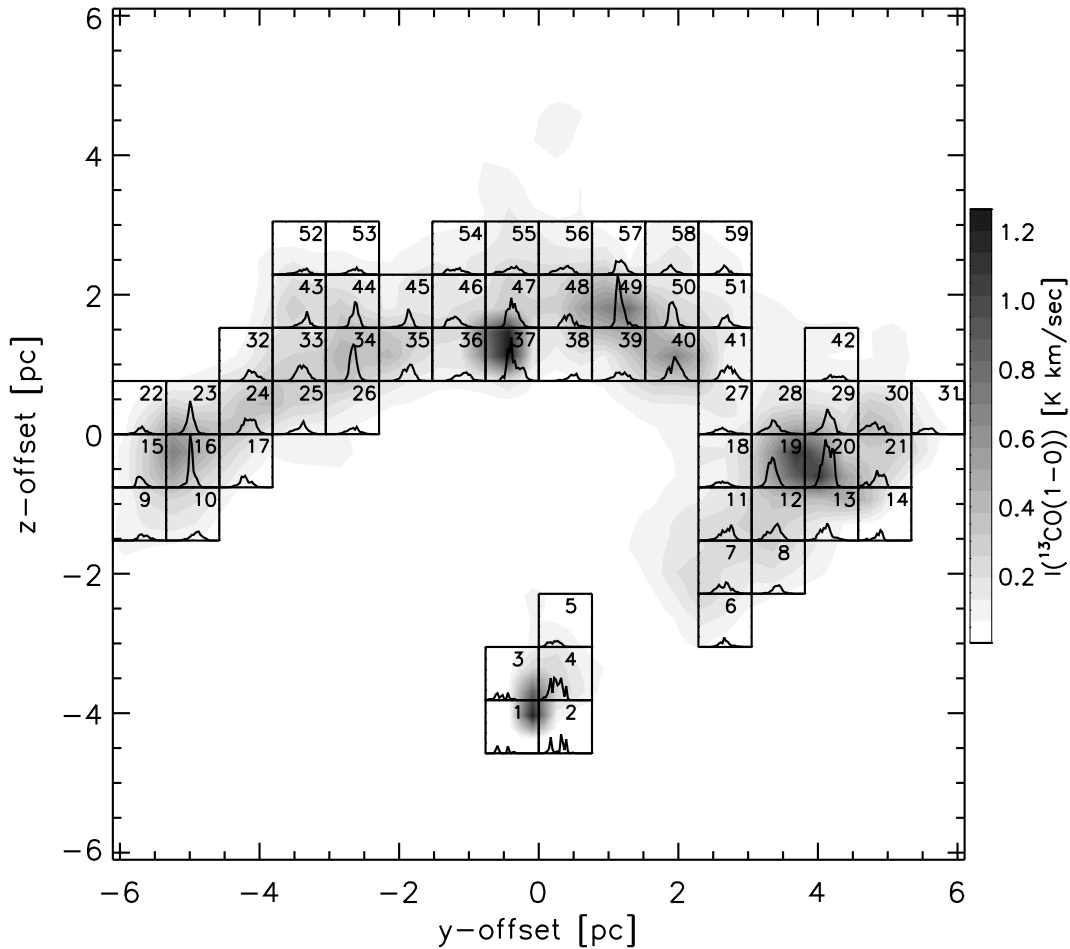}
  \end{center}
  \caption{\label{f:13COmapyz} Total intensity map (in K~km~s$^{-1}$)
    of $^{13}$CO emission from the model cloud, seen along the
    inflows. Line-of-sight velocity spectra are overplotted for
    regions with $I\geq 2.5$~K~km~s$^{-1}$.  The numbers refer to
    Table~\ref{t:histo}.}
\end{figure*}

Figure~\ref{f:13COmapyz} shows the velocity-integrated emission
(grayscale) of the $^{13}$CO-map for a model cloud assembled by
large-scale colliding flows (model Gf2 of \citealp{2008ApJ...674..316H}, at $t = 14.5$~Myr). The
view is along the inflow ($x$) direction, and the frames measure
$12$~pc across. Overplotted are the spectra for
regions with $T_b\geq 2.5$~K. Each spectrum covers a velocity range from
$-2$ to $+2 \kms$. Two features in the line profiles are noteworthy:
they are frequently asymetric, and some cases
they exhibit more than a single peak.  

Table~\ref{t:histo} lists the first moments of the line
profiles. Column 1 shows the number of the region, as shown in
Figure~\ref{f:13COmapyz}, and column 2 shows the intensity in
K~$\kms$. Columns 3 \& 4 list the centroid velocity and the velocity
dispersion, respectively, and finally, in column 5 we report whether
the line profile has multiple peaks or not.

\begin{figure*}
  \includegraphics[width=0.49\textwidth]{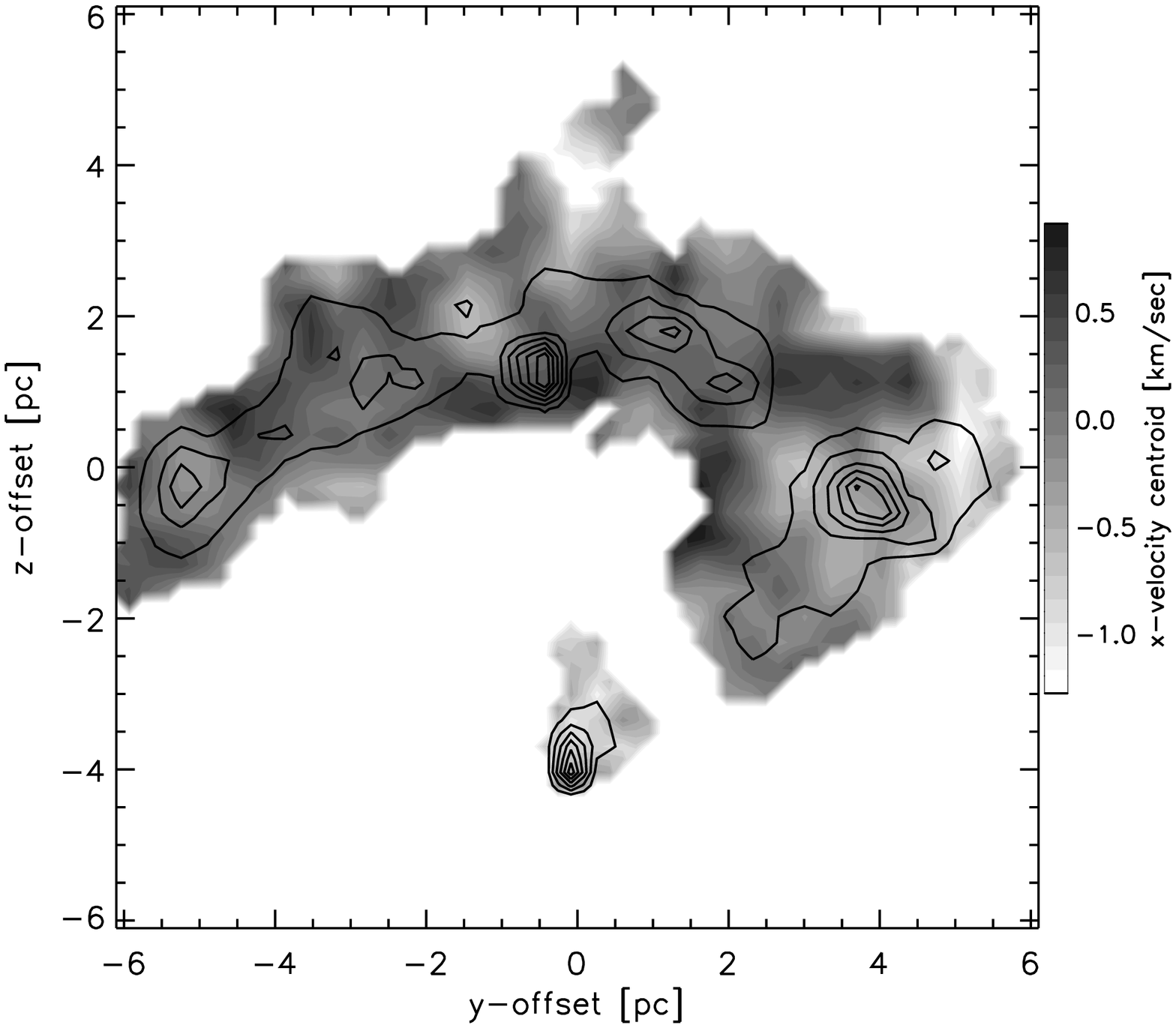}
  \hfill
  \includegraphics[width=0.49\textwidth]{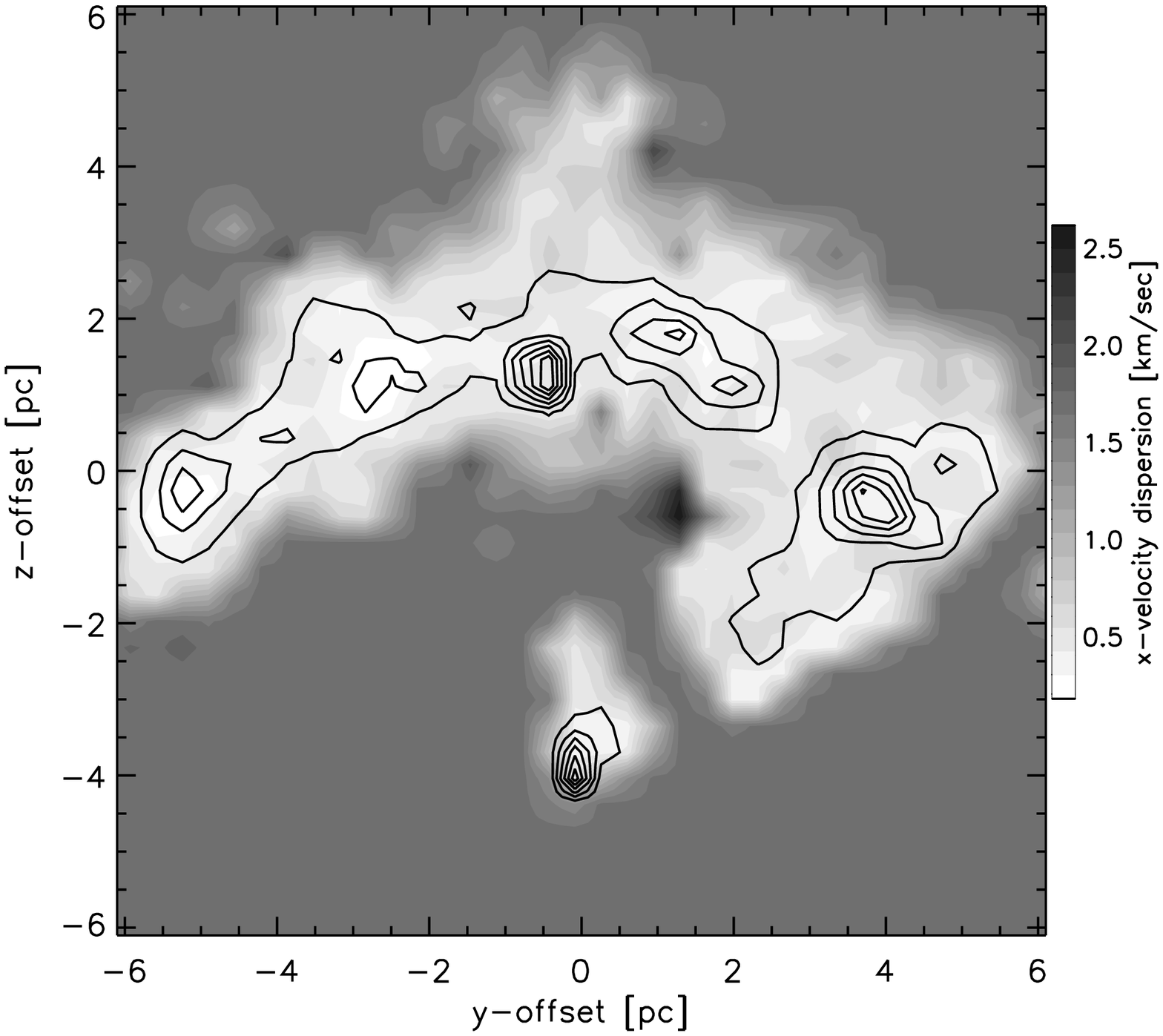}
  \caption{\label{f:13COcentsig} {\em Left:} Centroid velocity (in km~s$^{-1}$) of the
    model cloud, seen along the inflows. The filament is
    coherent in velocity space. {\em Right:} Velocity dispersion (in $\kms$) of the
    model cloud, as seen along the inflows.  The densest
    regions have the smallest velocity dispersion.}
\end{figure*}

Figure~\ref{f:13COcentsig} (left) shows the centroid
velocity and the velocity dispersion (right), again in the $y-z$ projection.
Despite the fact that we view the cloud along the inflow direction,
the dense filament is coherent in velocity space (see also centroid
velocities in center column of Table~\ref{t:histo}), except for the
largish core on its right end, which is approaching the observer at
$\approx -0.5$~km~s$^{-1}$. Thus, the velocity dispersion in the dense
gas -- the core-to-core velocity dispersion -- does not contain much
information about the velocities of the assembling flows ($\sim
7.9~\kms$), in fact, the motions in the (coherent) dense post-shock gas
are subsonic.

\begin{table}
{
\begin{center}
\begin{tabular}{rcccc} 
\hline
number & $I$ [K~km~s$^{-1}$] & $\langle v\rangle$ [km~s$^{-1}$] &
  $\sigma v$ [km~s$^{-1}$] & mult\\ 
\hline 
\hline
$ 1  $ & $  3.75$ & $  -0.65 $ & $ 0.82 $ & yes\\ 
$ 2  $ & $  9.31$ & $  -0.48 $ & $ 0.67 $ & yes\\ 
$ 3  $ & $  3.67$ & $  -0.78 $ & $ 0.72 $ & yes\\ 
$ 4  $ & $ 10.92$ & $  -0.67 $ & $ 0.51 $ & yes\\ 
$ 5  $ & $  2.94$ & $  -0.79 $ & $ 0.86 $ & yes\\ 
$ 6  $ & $  4.66$ & $   0.01 $ & $ 0.50 $ & no\\ 
$ 7  $ & $  5.59$ & $  -0.05 $ & $ 0.54 $ & yes\\ 
$ 8  $ & $  4.19$ & $  -0.10 $ & $ 0.43 $ & no\\ 
$ 9  $ & $  3.04$ & $   0.35 $ & $ 0.64 $ & no\\ 
$ 10 $ & $  4.41$ & $   0.25 $ & $ 0.57 $ & no\\ 
$ 11 $ & $  7.69$ & $   0.12 $ & $ 0.54 $ & yes\\ 
$ 12 $ & $  8.20$ & $  -0.30 $ & $ 0.50 $ &  yes\\ 
$ 13 $ & $  8.39$ & $  -0.41 $ & $ 0.53 $ & yes\\ 
$ 14 $ & $  5.07$ & $  -0.52 $ & $ 0.53 $ & no\\ 
$ 15 $ & $  5.20$ & $   0.11 $ & $ 0.39 $ & no\\ 
$ 16 $ & $ 25.99$ & $  -0.11 $ & $ 0.29 $ & no\\ 
$ 17 $ & $  5.67$ & $  -0.05 $ & $ 0.46 $ & yes\\ 
$ 18 $ & $  2.90$ & $  -0.19 $ & $ 0.61 $ & no\\ 
$ 19 $ & $ 14.46$ & $  -0.45 $ & $ 0.48 $ & no\\ 
$ 20 $ & $ 23.35$ & $  -0.31 $ & $ 0.43 $ & no \\ 
$ 21 $ & $  8.28$ & $  -0.56 $ & $ 0.53 $ & yes\\ 
$ 22 $ & $  3.77$ & $   0.13 $ & $ 0.48 $ & no \\ 
$ 23 $ & $ 16.29$ & $  -0.17 $ & $ 0.36 $ & no \\ 
$ 24 $ & $  7.53$ & $   0.27 $ & $ 0.56 $ & no \\ 
$ 25 $ & $  6.25$ & $   0.15 $ & $ 0.53 $ & no \\ 
$ 26 $ & $  3.60$ & $   0.05 $ & $ 0.51 $ & no\\ 
$ 27 $ & $  3.13$ & $  -0.15 $ & $ 0.63 $ & no\\ 
$ 28 $ & $  6.81$ & $  -0.34 $ & $ 0.61 $ &  no\\ 
$ 29 $ & $ 12.30$ & $  -0.25 $ & $ 0.47 $ & no \\ 
$ 30 $ & $  5.77$ & $  -0.85 $ & $ 0.58 $ & yes\\ 
$ 31 $ &  $ 2.81$ & $  -0.81 $ & $ 0.73 $ & no\\ 
$ 32 $ & $  5.29$ & $   0.46 $ & $ 0.71 $ & no \\ 
$ 33 $ & $  7.63$ & $   0.25 $ & $ 0.55 $ & no \\ 
$ 34 $ & $ 17.64$ & $   0.08 $ & $ 0.33 $ & no\\ 
$ 35 $ & $  7.99$ & $   0.22 $ & $ 0.52 $ & yes\\ 
$ 36 $ & $  4.26$ & $   0.31 $ & $ 0.68 $ & yes\\ 
$ 37 $ & $ 21.40$ & $   0.12 $ & $ 0.52 $ & yes\\
$ 38 $ & $  3.53$ & $   0.53 $ & $ 0.80 $ &  no\\ 
$ 39 $ & $  4.22$ & $   0.13 $ & $ 0.67 $ & no\\ 
$ 40 $ & $ 11.92$ & $   0.25 $ & $ 0.55 $ & no\\ 
$ 41 $ & $  7.41$ & $   0.24 $ & $ 0.59 $ & no\\ 
$ 42 $ & $  2.99$ & $   0.27 $ & $ 0.65 $ & no\\ 
$ 43 $ & $  7.96$ & $   0.45 $ & $ 0.68 $ & no\\ 
$ 44 $ & $ 12.80$ & $   0.19 $ & $ 0.41 $ & no\\ 
$ 45 $ & $  9.20$ & $   0.16 $ & $ 0.45 $ & no\\ 
$ 46 $ & $  5.43$ & $  -0.29 $ & $ 0.53 $ & no\\ 
$ 47 $ & $ 14.63$ & $   0.08 $ & $ 0.45 $ & no\\ 
$ 48 $ & $  6.65$ & $   0.28 $ & $ 0.60 $ & yes\\ 
$ 49 $ & $ 25.58$ & $   0.14 $ & $ 0.43 $ &  no\\ 
$ 50 $ & $ 12.54$ & $   0.09 $ & $ 0.36 $ & no\\ 
$ 51 $ & $  6.06$ & $   0.15 $ & $ 0.52 $ & no\\ 
$ 52 $ & $  2.95$ & $   0.19 $ & $ 0.66 $ & no\\ 
$ 53 $ & $  3.48$ & $   0.14 $ & $ 0.58 $ & no\\ 
$ 54 $ & $  3.27$ & $  -0.19 $ & $ 0.61 $ & no\\ 
$ 55 $ & $  4.05$ & $  -0.06 $ & $ 0.63 $ & no\\ 
$ 56 $ & $  4.25$ & $  -0.09 $ & $ 0.56 $ & no\\ 
$ 57 $ & $  6.97$ & $   0.18 $ & $ 0.50 $ & no\\ 
$ 58 $ & $  4.67$ & $  -0.05 $ & $ 0.47 $ & no\\ 
$ 59 $ & $  4.47$ & $  -0.05 $ & $ 0.46 $ & no\\ 
\hline
\end{tabular}
\end{center}
}
\caption{\label{t:histo}Parameters of subfields in
  Figure~\ref{f:13COmapyz}. The last column indicates the existence of
  an obvious second component.}
\end{table}

\begin{figure}
  \begin{center}
    \includegraphics[width=0.8\columnwidth]{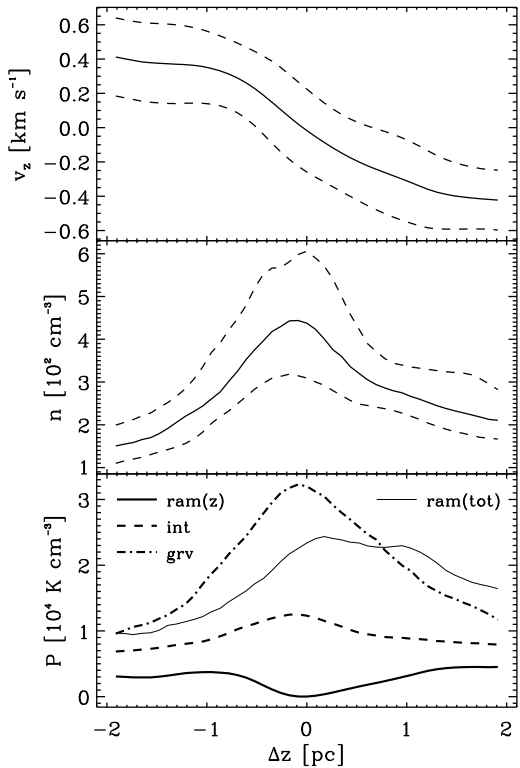}
  \end{center}
  \caption{\label{f:vprofile} Velocity, density and pressure profiles
    of gas falling onto the filament, in the plane of sky, against
    distance to the filament. Negative distances are located below the
    filament. For the velocities, the solid line denotes the centroid
    (mean) velocity, and the dashed lines indicate the $1\sigma$
    scatter. For the density, solid lines stand for median values,
    while dashed lines indicate the lower and upper quartiles. The
    pressure plot shows the ram, internal and gravitational pressure
    as indicated. The thin solid line stands for the total ram
    pressure, i.e. all three components of the velocity contribute.}
\end{figure}

This also can be seen when considering the gas motions perpendicular
to the filament and in the plane of the sky, i.e. the infall of
more diffuse gas onto the filament (Fig.~\ref{f:vprofile}). 
The velocity profiles were calculated by identifying the filament
in the two-dimensional projection (see Fig.~\ref{f:13COmapyz}).
Comparing the velocity and the density profiles (top and
center panel), it is clear that gas at higher densities should show
smaller velocity dispersions: at a distance of $2$~pc from the axis of
the filament, the infall velocity ranges between $0.2$ and
$0.6$~km~s$^{-1}$, i.e from subsonic to slightly supersonic. In other 
words, the dense filament is essentially a post-shock region
(with subsonic internal motions), confined by a dynamical pressure
generated by the (supersonic) infall of gas onto the filament.
We note that the dense filament is not very well resolved and that
thus we decided not to model higher-density tracers showing the subsonic
motions more clearly -- the line-profiles would not be reliable. Our
choice of $^{13}$CO probes the immediate vicinity of dense cores, which
makes it an appropriate tracer for our purposes. 

The formation of subsonic cores in post-shock gas has also been discussed
in the context of 
driven, supersonic turbulence assembling dense cores 
\citep{2005ApJ...620..786K,2007prpl.conf...63B}. In that case, density
peaks form at locations of maximum compression and of minimum relative
velocity difference. In our case (Fig.~\ref{f:vprofile}), the inflows
are driven by the deepening gravitational potential of the assembling cloud,
not relying on externally forced turbulence.

The bottom panel of Figure~\ref{f:vprofile} compares the median value
of the pressures as an estimate of what processes dominate the
gas dynamics in the environment of the filament.
The thick solid line corresponds to the ram pressure obtained with the
infall velocity along the $z$-direction (i.e. perpendicular to the
filament in Figure~\ref{f:13COmapyz}), while the thin solid line
shows the total ram pressure profile.  The gravitational pressure $\rho\Phi$ 
is shown by the dot-dashed line, while
the thermal pressure is given by the dashed line.  Clearly, the region
is more dynamical than just indicated by the mean infall velocities:
as suggested by this figure, the infall (upper panel) is driven by the
gravitational potential (lower panel), while the thermal pressure
responds only weakly to the gravitational pressure.

\begin{figure*}
  \begin{center}
  \includegraphics[width=0.85\textwidth]{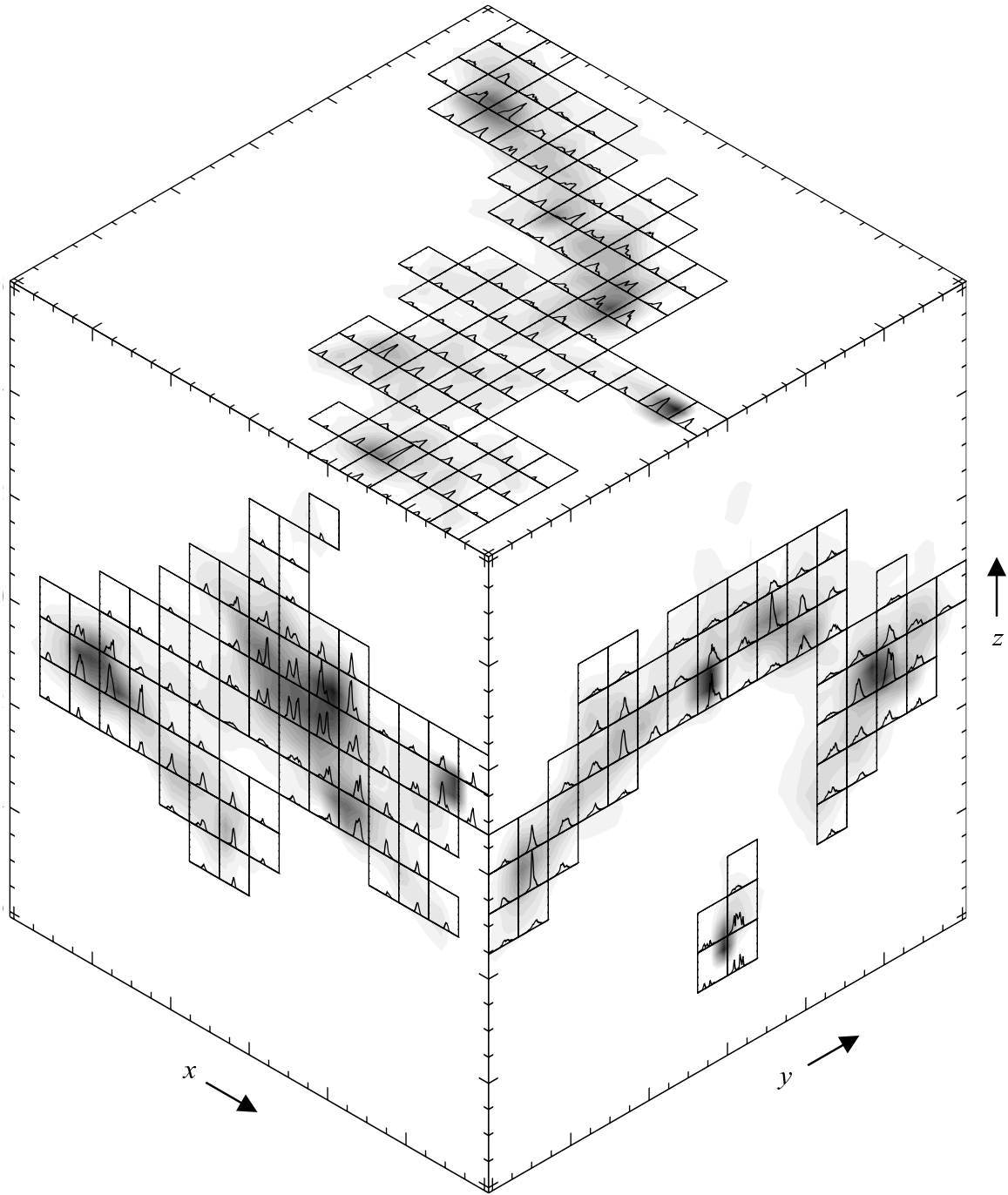}
  \end{center}
  \caption{\label{f:cube} The three intensity projections of the datacube.
          The flows are along the $x$-axis, and Figure~\ref{f:13COmapyz} 
          shows the $(y,z)$-plane.}
\end{figure*}

To get a clearer view of the dynamics, 
we show in Fig.~\ref{f:cube} the
three projections of the datacube, with the line profiles overplotted,
as in Fig.~\ref{f:13COmapyz}. The $x-y$ panel is on the top of the
cube, the $x-z$ on the left, and the $y-z$ on the right. These
projections reveal that the filament is actually tilted by approximately
$45$ degrees in the $x-y$ plane, with the left side closer to the
observer located at $x$~$\to$~$\infty$. From the 
three-dimensonal datacube, we know that the infall velocities in the
three directions are of similar magnitude. However, this does not
necessarily translates to an infall, double-peaked line
profile. Although the three projections exhibit some double-peaked
profiles, most of our $^{13}$CO lines exhibit only asymmetric,
supersonic ($\sigma_v\sim 0.6 \kms$) profiles. Thus, the supersonic
line widths observed in molecular clouds that are usually attributed
to {\em random} motions (or "turbulence"), very likely contain a substantial 
component of {\em ordered}
motions due to coordinated global collapse.
A filament tilted with respect to the line-of-sight -- a very likely scenario --, 
only helps to suppress clear signatures of coordinated infall onto the filament,
giving the impression of "turbulent" motions.

Table~\ref{t:filament} summarizes the measured properties of the model
filament, as identified in the two-dimensional projection.  
Since the massive cores potentially could affect the
geometry, we show numbers for the full filament (including the
cores), and the left half of the filament, which does not contain
massive cores.

\begin{table}
\begin{center}
\begin{tabular}{lccccc}
\hline
              & $M$         & $T$  & $L$  & $m_c$      & $m_c$  \\
              & [M$_\odot$] & [K]  & [pc] & measured & expected \\       
\hline
\hline
full filament & $1.1\times 10^3$ & $28$ & $12$ & $9.0$ & $4.5$ \\
without cores & $3.5\times 10^2$ & $27$ & $5.7$& $6.2$ & $4.4$ \\
\hline
\end{tabular}
\end{center}
\caption{\label{t:filament}Filament parameters for models, to be
  compared to the toy model of \S\ref{s:toymodel}. The mass per length
  $m_c$ is given in M$_\odot$~pc$^{-1}$.  The first line gives the
  numbers for the full filament (as used for Figure~\ref{f:vprofile}),
  while the second line refers to the left half of the filament which
  does not contain any cores.}
\end{table}

We note that the filament line density is approximately twice the
expected limit for the isothermal case (for the higher temperature of
28~K in the simulation).  As mentioned above, the filament is tilted,
resulting in a shorter projected length, and thus in a higher line density.
Also, other effects such as limited resolution and the possibility that the
filament is not static but is still accreting mass on timescales of order $1$~Myr,
could play a role.

Given that the model filament does not agree with the static approximation,
we use equation (\ref{eq:vm}) for the full filament to estimate $v
\sim 0.4 {\ln (\ro/R)}^{1/2} \kms$.  What value of $\ro$ to use is not
clear, but if we take $\ro = 2$~pc and $R = 0.2$~pc for typical
values, $v \sim 0.6 \kms$, this is about twice the average infall
velocity at the filament (Figure~\ref{f:vprofile}).

To summarize, 
Figure~\ref{f:vprofile} demonstrates the overall collapse of the cloud in
addition to acceleration onto the filament. Obviously, equation
(\ref{eq:deltav}) can describe the gas dynamics of the simulated cloud only 
in a very general way, as (a) matter is being added to the cloud during its
collapse from the inflows, (b) the cloud (or filament) cross-section is not
exactly circular, and as (c) there are shear components in the flows. 
Nevertheless, the overall collapse motions are approximately cylindrical
(see total ram pressure profile in Fig.~\ref{f:vprofile}), indicating that
at this stage of the simulation the massive filament dominates the gravitational
potential and thus the dynamics of the cloud, 
and justifying equation~(\ref{eq:deltav}) as a rough description
of the actual dynamics. The total mass of the cloud,
including the filament itself, at this epoch is $\sim 2000
\msun$ distributed over a region approximately $10 \times 6$~pc, which
implies an average surface density of $\Sigma \sim 7 \times 10^{-3}$~g~cm$^{-2}$
or $1.8\times 10^{21}$~cm$^{-2}$.
Assuming the initial radius in the $y$-direction was $\sim
10$~pc, equation (\ref{eq:deltav}) then implies a velocity gradient
across a region of $2$~pc of $\delta v \sim 0.4 \kms$, which again is in
reasonable agreement with the results in Figure 2.  Considering the
very crude nature of these estimates, agreement with the numerical
simulation at the factor of two level is adequate.


\section{Comparison with the Pipe Nebula}\label{s:pipecomp}

Motivated by the global, approximately cylindrical infall observed in the
simulation (\S\ref{s:numresults}) under rather general conditions, we 
first discuss in this section our analytical estimates (\S\ref{s:toymodel}) and then
the simulation results in the context of the Pipe nebula. We summarize
the relevant observational findings first, and then use the approximate
agreement between observations and models to interpret the
physical conditions of the Pipe nebula in the context of global infall
(\S\ref{s:discussion}).

The most recent estimate of core properties comes from
\citet{2009arXiv0904.4169R}.  The core masses range from $\sim 0.3 \msun$
to about $25 \msun$, with radii ranging from $\sim 0.04$ to $\sim
0.2$~pc and a typical internal density (H$_2$) of $\sim 10^4~\cm3$.
The FWHM of the C$^{18}$O lines ranges from about 0.4 to about $1
\kms$.  The estimated visual extinction rises slowly with mass until
about $M \sim 2 \msun$, at which point $A_V$ rises rapidly from about
$\sim 4$ to $10 - 15$; this is roughly the mass range which becomes
gravitationally bound, according to \citet{2008ApJ...672..410L}.  Median
properties of the cores are $M \sim 0.8 \msun$, $R \sim 0.07$~pc, and
FWHM $\sim 0.36 \kms$.  Such a ``median'' core is typically
pressure-confined according to \citet{2008ApJ...672..410L}.  To simplify the
comparison with observations, we assume a typical temperature of 10 K,
though there is some evidence that some of the Pipe cores have
slightly higher temperatures, $\sim 12 - 15$~K
\citep{2007ApJ...662.1082R}.

To proceed further we need an estimate of the average line density of
the filament.  The Pipe, though highly elongated, is not a perfectly
straight filament nor is it uniform.  The sum of the core masses in
\citet{2009arXiv0904.4169R} is $228 \msun$; distributing this mass
over a length $\sim 9^{\circ} \sim 20$~pc results in an average line
density $\sim 11 \msun {\rm pc}^{-1}$.  This estimate neglects the
inter-core mass but includes the ``bowl'' region which is much more
complex in structure.  If we restrict attention to the portion of the
pipe at negative galactic longitudes, the total core mass is $105
\msun$ over $\sim 15$~pc or $m \sim 7 \msun {\rm pc}^{-1}$.  As the
Pipe does not exhibit considerable star formation at the present
epoch (\citet{2009arXiv0908.4086F} estimate a star formation efficiency of $0.06$\% when
comparing to the total cloud mass given by \citealp{1999PASJ...51..871O}), 
it seems reasonable to assume an average line density of $m
\sim 15 \msun {\rm pc}^{-1}$, to include mass outside the core but not
so much that the filament would be radially gravitationally-unstable.

To further fix ideas we take an average filament optical depth to be
$A_V \sim 2$.  This is near the low end of the core extinction values
found in \citet{2009arXiv0904.4169R}.  Then from
equation (\ref{eq:h}), the scale height is $H \sim 0.095$~pc; this
seems reasonable, seeing that the median core radius given by
\citet{2009arXiv0904.4169R} is about 0.07 pc.  We may also then derive a
central density from equation (\ref{eq:hrho}) of $\rho_0 \sim 9.5
\times 10^{-21} \gcm3$ or $n(H_2)_0 \sim 2.4 \times 10^3 \cm3$.  This
density is a few times smaller than the average core density $n(H_2)
\sim 10^4 \cm3$, but we have not taken any confining pressure into
account.

We are now in a position to calculate an estimate of the infall
velocity using equation (\ref{eq:vin}).  Taking $\ro/R = 10$, we find
$v \sim 0.77 \kms$.  The timescale for this motion is sensitive to the
choice of initial condition, i.e. the initial velocity; for the above
solution, the time taken to fall in from $0.9$ to $0.1$~pc is about $2$~Myr.

The translation of this infall velocity into a typical velocity
dispersion is sensitive to geometry at the factor of two level.
Assuming that the filament is oriented perpendicularly to our line of
sight, and the emission is simply mass weighted, the average velocity
width is $\sim (4/\pi) \times 0.77 \kms$.  If we identify this with
the FWHM or line width, our results are consistent with the median
$^{13}$CO line width of $\sim 1.2 \kms$ found by \citet[][see their
  Fig.~6b]{1999PASJ...51..871O}.  It is larger than that observed in the
C$^{18}$O cores by about a factor of $2.5$; however, as pointed out by
\citet{2008ApJ...672..410L}, the C$^{18}$O observations sample material both
outside and inside the cores.  As mentioned in \S\S 2 and 3, cores
would be predicted to be the sites of post-shock gas and thus with
much smaller internal velocity dispersions.  Lada \etal state that, in
the lower-mass cores, half or more of the total line-of-sight column
density arises from material outside the core (also 
\citealp{2007ApJ...669.1042G,2009ApJ...699..230G}). 
Conversely, this means
that for many cores, half or less of the emitting material is outside
the core.  Thus we would predict a core velocity width perhaps half
that above, or around $\sim 0.4 \kms$, in rough agreement with
observations \citep{2007ApJ...671.1820M}.  In addition, there is
no specific reason to assume that the filament is oriented perpendicularly
to the line of sight. Any (very likely) tilt of the filament with respect
to the line of sight will result in an even smaller projected velocity
dispersion.

Going back to the numerical simulations, we note that the velocity
dispersion in the flow direction, calculated as the second moment of
the line profile (see Fig.~\ref{f:13COmapyz} and Table~\ref{t:histo}),
are or the order of $\sim 0.6 \kms$. This implies that the FWHM,
calculated as $\sqrt{8 \ln{2}} \sigma_v$, is of the order of $\sim 1.4
\kms$, a value close to the observed $1.2 \kms$ by
\citet{1999PASJ...51..871O} for the Pipe.  The discrepancy with the analytic
result, however, is probably due in part to the more complex velocity
field of the simulation, including the global collapse.  In some cases
the observed line width is inflated because of the superposition of
separate cores along the line of sight, which can be seen in systems
with two peaks of emission (see also last column in
Table~\ref{t:histo}).  In any case the simulations are in reasonable
agreement with the Onishi \etal observations.

While there are other possible mechanisms to generate the observed
velocity dispersions in molecular gas (one of them would be
MHD turbulence, see e.g. \citet{1995ApJ...440..686M} for a detailed
discussion), 
we point out that in a cloud of many Jeans masses, of non-uniform density
and of irregular geometry, it will be impossible to avoid having significant
gravitational accelerations locally. These in turn will produce motions
on the order of $1$~km~s$^{-1}$ (see Fig.~3 and \S\ref{s:toymodel}),
unless there is a special forcing or velocity field that would prevent
this. Yet this forcing not only would have to be chosen so cleverly in
time and space as to support just the right region that is about to collapse, but it also
would have to prevent the collapse while reproducing the supersonic
motions in the environment of the filament. Thus while we cannot say that
there is no other source for the observed gas motions, we do argue that at
least part of these motions are likely to be gravity-driven.

\section{Discussion \& Conclusions}\label{s:discussion}

The results presented in this work support the idea that molecular
clouds are in a general state of global collapse, suggested more than
35 years ago by \citet{1974ApJ...189..441G}.  Although such global
collapse of irregular structures develops internal turbulence at some 
level, the large linewidths in MCs are caused mainly by
the large-scale systematic inward motions in this scenario.
Given the irregularities, angular momentum conservation
and the high Reynolds numbers, it would actually be outright surprising
if the collapse did {\em not} generate some "turbulence".

\citet{1974ApJ...192L.149Z} suggested that such large-scale collapse would
result in a star formation rate much higher than observed, mandating some
mechanism of cloud support to render star formation inefficient
(see \citealp{1999osps.conf...29M} for a summary). Yet, analytical
and numerical studies \citep[e.g.,][]{2004ApJ...616..288B,
2008MNRAS.385..181F,2008ApJ...674..316H,2008ApJ...689..290H,
2007ApJ...657..870V,2008MNRAS.390..769V,2009arXiv0904.4515V}
demonstrate that MCs, while globally collapsing, are highly susceptible to strong fragmentation. 
While this fragmentation is to some extent a consequence
of a combination of thermal and dynamical instabilities during
the cloud formation process \citep{2008ApJ...683..786H}, 
it is mainly due to the fact that the cloud geometry is finite,
leading to gravitational edge focusing, i.e. non-linear gravitational
accelerations as a function of position \citep{2004ApJ...616..288B}.
In other words, the existence of non-linear
gravitational accelerations as a function of position, leading to
a rapid piling up of material as well as local fragmentation, allows
local collapse to proceed faster than global collapse. And it is
precisely because small, high-density structures are rapidly developing, that
the mass involved in the densest regions has to be small, ensuring a
small star formation efficiency. In other words, the observed low star
formation efficiency seems to be the testimony of the importance of
non-linear acceleration at particular places while the global collapse
occurs. 
In this connection it is worth pointing out that the major region of star formation
within the Pipe (with 15 YSOs; \citealp{2009AAS...21333705C}), the B59 cloud, lies at one 
end of the filament, which is a preferred locus for gravitational edge focusing
\citep{2004ApJ...616..288B}.

While there are clearly a number of uncertainties in both the analytical
and numerical calculations and in the precise observational
quantities, a simple model of gravitational infall toward the filament
clearly can account for the observed non-thermal velocity dispersions
in the Pipe.  The gravity-driven model also does 
predict small (core-to-core) velocity dispersions consistent with observations
\citep{2002ApJ...578..914H,2004ApJ...614..194W,2007ApJ...655..958W,2009arXiv0904.4511K}.
This agreement is possible because the ``turbulence'' is not being
continuously driven by an ad hoc force but is the result of
(global)  gravitational acceleration. The simulation results also suggest
that while gas is falling onto the filament supersonically, the velocity 
dispersions in the dense post-shock gas are subsonic, consistent with
the more detailed treatment of core formation in post-shock gas by
\citet{2007ApJ...669.1042G,2009ApJ...699..230G}. Yet the current simulation is not
sufficiently resolved to support more detailed statements about the
dense cores.

\citet{2008ApJ...672..410L} suggested that the confining pressure for many
Pipe cores was the result of the ``weight'' of the cloud.  Here we
refine this suggestion by pointing out that a static pressure estimate
is not strictly appropriate for a cloud in supersonic motion.  The ram
pressure that we infer is, however, another aspect of the same
physical mechanism - gravitational acceleration.  It is also worth
noting that, in this interpretation, the non-thermal velocity
dispersion in the cores and surrounding regions are dominated by
collapse motions and thus do not provide pressure support against
gravity \citep{1999ApJ...515..286B,2006MNRAS.372..443B,2008MNRAS.390..769V}. 
Nor does in this picture a driving source for turbulence
within molecular clouds appear to be necessary \citep[see
also][]{2008MNRAS.385..181F,2008ApJ...689..290H}.

The numerical simulations, with an average infall velocity in the
$y$-direction of $\sim 0.4 \kms$ and a density of $\sim 10^3 \cm3$ at
the filament, imply ram pressures of order $P/k \sim 2 \times 10^4
\cm3$~K.  
Scaling this result from the $\sim 1000
\msun$ filament of the simulation to the $\sim 3000 \msun$ estimated
for the Pipe region from $^{13}$CO \citep{1999PASJ...51..871O} results in a
predicted pressure 3 times larger, or about $6 \times 10^4 \cm3$~K.
This is in reasonable agreement with the confining pressures for Pipe
cores estimated by \citet{2008ApJ...672..410L} of $P/k \sim 8 \times 10^{4}
\cm3$~K.

The core mass function (CMF) inferred for the Pipe 
(\citealp{2007A&A...462L..17A}; \citealp{2008ApJ...672..410L};
\citealp{2009arXiv0904.4169R}) peaks at a mass well above the
typically-estimated peak of the stellar initial mass function (IMF).
In the gravity-driven picture, translating the CMF at an instant of
time to an IMF is complicated because the model of
gravitationally-accelerated infall implies that the filament - and
thus the cores - are accreting mass (see \citealp{2007MNRAS.379...57C} 
for a related argument). If we set the ram pressure of
infall equal to confining pressure inferred by \citet{2008ApJ...672..410L},
and use $v = 0.77 \kms$ from the analytic model, the implied density
of infalling material $\rho \sim 1.9 \times 10^{-21} \gcm3$ or $N(H_2)
\sim 480 \cm3$.  The mass accumulation rate over a cylindrical region
of radius $R = 0.1$~pc and length $z = 0.2$~pc is then $2 \pi R z \rho
v \sim 2.7 \times 10^{-6} \msunyr$. 
While this estimate is obviously sensitive to the adopted
parameters, it is clear that the lower-mass cores can quite plausibly
double their mass over timescales of less than 1 Myr \citep[see
also][]{2007ApJ...669.1042G,2009ApJ...699..230G}.  In fact, observations of starless cores are
starting to indicate that quiescent, coherent dense starless cores do
accrete actively \citep[e.g.,][]{2007ApJ...671.1839S}.  Alternatively, the
numerical simulation shows a doubling of mass in cores over a
timescale of 1 Myr \citep{2008ApJ...689..290H}, consistent with this estimate.

The long-range nature
of gravity makes it very difficult to avoid the generation of
supersonic velocity fields and the production of filaments and other
dense structures in clouds with many Jeans masses
\citep{2004ApJ...616..288B,2002MNRAS.332L..65B,2003MNRAS.339..577B,2007ApJ...657..870V,
2008ApJ...674..316H,2009MNRAS.tmp..667S}.

The mass, spatial extent, and typical column densities of the Pipe are
not very different from those of the Taurus molecular cloud.  It is
conceivable that the Pipe is representative of how Taurus would have
appeared 1 - 2 Myr ago.  The average filament line density in Taurus
is estimated to be roughly twice the critical value
\citep{2002ApJ...578..914H}, which is consistent with the active star formation
in the region.  Blindly applying the mass infall rate calculated above
to Taurus would imply a buildup from $\sim 15 \msun {\rm pc^{-1}}$ to
the observed $\sim 30 \msun {\rm pc^{-1}}$ on a timescale of 1 Myr.

Finally, one may consider the future evolution of the Pipe nebula with regards
to the star formation efficiency.
While we have shown that besides the gas dynamics, the gravity-driven model can 
naturally explain the origin of the strong fragmentation as a key ingredient
for the observed low star formation efficiency, we have not discussed
how the remaining molecular gas can be prevented from eventually collapsing
onto the dense filament. In low-mass star-forming regions this could be
achieved by e.g. magnetic support of the diffuse molecular gas 
\citep{2008ApJ...680..420H,2009arXiv0904.4071P}, or by Galactic tidal disruption 
\citep{2009MNRAS.395L..81B,2009MNRAS.393.1563B}.

\acknowledgments
We thank the referee for comments and questions that 
helped to improve our presentation.
This work was supported in part by NSF grant AST-0807305
and by the University of Michigan. Computations were
performed at the National Center for Supercomputing 
Applications (AST 060034). J.P.-B. acknowledges support
from grant UNAM-PAPIIT IN110409.
This work has made use of 
NASA's Astrophysics Data System.

%
%
\bibliographystyle{apj}
\bibliography{./references}


\end{document}